\documentclass[prb,nopacs,twocolumn]{revtex4}

\usepackage[dvips]{graphicx}
\usepackage{dcolumn}
\usepackage{bm}
\def\be{\begin{equation}}
\def\en{\end{equation}}

\def\p{\partial} 
\newcommand{\av}[1]{\langle{#1}\rangle}
\def\gs{\gtrsim}

\newcommand{\bi}[1]{\mbox{\boldmath$#1$}}
\def\p{\partial}
\def\bea{\begin{eqnarray}}
\def\ena{\end{eqnarray}}

\def\ve{\varepsilon}

\begin{document}

\title{Solvation and Dissociation  in Weakly Ionized 
Polyelectrolytes}

\author{Akira Onuki and Ryuichi Okamoto}
\affiliation{Department of Physics, Kyoto University, Kyoto 606-8502,
Japan}
\date{\today}

\begin{abstract}
We present a Ginzburg-Landau theory of 
inhomogeneous  polyelectrolytes 
with a polar solvent. 
First, we take into account the molecular 
(solvation) interaction   among the ions, the charged 
monomers,  the uncharged monomers, and the 
solvent molecules,   together with the electrostatic 
interaction with a composition-dependent dielectric 
constant.  Second, we treat 
the degree of ionization as  a fluctuating  variable  
 dependent  on the local electric potential.  
With these two ingredients included,  
our  results are as follows. 
(i)  We derive a mass reaction law and 
a general expression for the surface tension.  (ii) We calculate  
the structure factor of the composition 
fluctuations as a function of various parameters 
of the molecular interactions, which provides 
 a general criterion  of the formation of 
 mesophases. 
(iii) We numerically examine some typical examples 
of interfaces and mesophase structures, 
which strongly depend on the molecular interaction parameters.   
\end{abstract}


\maketitle


\section{Introduction}

Polyelectrolytes are much more complex 
than low-molecular-weight 
electrolytes and neutral polymers \cite{PG,Barrat,Rubinstein}. 
Above all, the electrostatic interaction  
among the ionized monomers on the 
 polymer chains and the mobile ions 
strongly influence  the chain conformations and the 
mesophase formation \cite{Barrat,Rubinstein,Yoshikawa}. 
 Second, the dissociation (or ionization) on the chains 
should be treated as a chemical 
reaction in many polyelectrolytes 
containing weak acidic  monomers \cite{Joanny,Bu1,Bu2}, 
which is under the influence of 
the local electric potential. 
Then the degree of ionization  is 
a space-dependent annealed 
variable in inhomogeneous polyelectrolytes, 
while it has  mostly been treated  to be a given 
constant in the theoretical literature.  
Such ionization 
heterogeneity should not be negligible in structure formations 
and in phase separation. 
Third, for water-like solvents with large dielectric constant $\ve$, 
 polymers are often  hydrophobic and 
small ions are   hydrophilic \cite{Bu1}, which can also  
affect the phase behavior. 
However, not enough attention has yet been paid on 
the effects of such  short-range  molecular  interactions, 
where particularly relevant is the solvation (ion-dipole) 
interaction between ions and polar molecules \cite{Is}.

In this  paper we  hence 
treat the degree of ionization as a fluctuating 
variable and include 
the molecular interactions.  We  show their relevance 
in polyelectrolytes  
in  the simplest theoretical scheme. 
That is,  we use  the so-called random phase 
approximation \cite{PG}  
in the  Flory-Huggins scheme  of 
weakly charged polyelectrolytes  
\cite{Lu1,Lu2,PD,Krama}. 
On the basis of    a  
recent Ginzburg-Landau theory of ion distributions  
in binary mixtures 
\cite{Onuki-Kitamura,OnukiPRE,OnukiEPL,OnukiW},   
we account for the  solvation (hydration in aqueous solutions) 
between the  charged particles (ions and ionized monomers) 
and the solvent molecules, whose 
free energy contribution  usually 
much exceeds  the thermal energy 
$T$ (per charged particle)  \cite{Is}. 
Hereafter we set the Boltzmann constant equal to unity.

In one-phase states of weakly charged 
 polyelectrolytes, the structure factor 
 $S(q)$   of the composition fluctuations 
with wave number $q$ was  calculated 
in the  random phase approximation \cite{Lu1,Lu2}, where 
the solvation interaction was neglected. 
Let $\phi$ and $f$ be the polymer volume fraction 
and the fraction of charged monomers on the chains, respectively.  
In this approximation  the inverse of $S(q)$ is expressed as  
\be 
\frac{1}{S(q)}
={\bar r}(\phi) + \frac{q^2 }{12a\phi(1-\phi)} 
+ \frac{4\pi \ell_Bf^2}{v_0^2(q^2+\kappa^{2}) },
\en  
where the first term $\bar r$  depends  on $\phi$ and $T$,  
$v_0=a^3$ is the volume of a monomer, $\ell_B=e^2/\epsilon T$ 
is the Bjerrum length, and 
$\kappa$ is the Debye-H$\ddot{\rm u}$ckel 
wave number.  
Due to the last electrostatic term in Eq.(1.1), 
 $S(q)$ can have a peak  
 at an intermediate wave number $q_m$ for small $\kappa$ or for low salt 
concentration. We here mention some related 
experiments. (i) Such a peak 
has been observed 
by scattering  in one phase states of charged polymer systems 
\cite{Nishida,Shibayama}.  It indicates  
formation of mesophases in sufficiently poor 
solvent,  as was  
confirmed for a semidilue polyelectrolyte  solution 
\cite{Candau}. (ii) On the other hand, 
for  neutral, polar  binary mixtures\cite{Onuki-Kitamura,OnukiPRE} 
(or a mixture of neutral polymers 
and a  polar solvent \cite{OnukiW})
  with   salt near the critical point, 
we calculated  $S(q)$
 in the same form as  in Eq.(1.1).   In electrolytes,  
 the solvation  interaction 
 can  strongly affect  the composition 
fluctuations  particularly near the critical point. 
In fact, in a recent  scattering 
 experiment \cite{Seto}, 
a peak at an intermediate wave number 
  has been observed  in $S(q)$ in a near-critical binary mixture with  salt.  
(iii) We  also mention a finding of  
a broad peak in $S(q)$ 
in  semi-dilute solutions of 
neutral polymers in  a polar solvent 
with  salt \cite{Hakim,Hakim1}, 
where the solvation effect should be crucial. 
Thus we should calculate $S(q)$ 
in weakly charged polyelectrolytes 
including the solvation interaction.

We also mention calculations of the interface profiles 
in weakly charged polyelectrolytes in a poor solvent 
using the self-consistent field theory 
\cite{Shi,Taniguchi}. 
In these papers, however, the solvation 
interaction was neglected.  
In polyelectrolytes,
 the solvation 
interaction should decisively 
determine the charge distributions around an interface, 
 as in electrolytes \cite{OnukiPRE,OnukiW}.
 In addition,    the degree of ionization should significantly vary 
across an  interface in polyelectrolytes, because 
the dissociation process strongly depends on 
the local counterion density.

The organization of this paper is as follows.  
In Section 2, 
we will present a  Ginzburg-Landau approach accounting for 
 the molecular interactions  and the 
dissociation process. We will 
introduce the grand potential and present a theoretical 
expression for  the surface tension. 
In Section 3, we will calculate  
the composition structure factor 
generalizing the results in 
 the previous theories  \cite{Onuki-Kitamura,OnukiPRE,Lu1,Lu2}. 
In Section 4, we will 
numerically  examine the ion 
distributions around interfaces 
and in a periodic state.

\setcounter{equation}{0}
\section{Theoretical background} 

We suppose  weakly charged polymers  in a theta or poor 
solvent consisting of a one-component polar  fluid. We assume   
$f\ll 1$ to ensure flexibility of the chains. 
As suggested by Borue and Erukimovich\cite{Lu1} 
and by Joanny and Leibler \cite{Lu2}, 
the  random phase approximation can be used 
in concentrated solutions with  
\be 
\phi> f^{2/3}(\ell_B/a)^{1/3}.
\en 
We consider the semidilute case $\phi>N^{-1/2}$,  where 
$N$ is the polymerization index. Then 
the polymers consist of blobs with 
monomer number $g=\phi^{-2}$ and length $\xi_b = ag^{1/2}$ 
in the scaling theory \cite{PG} 
and the electrostatic energy within a blob 
is estimated as 
\be 
\epsilon_b= T(fg)^2\ell_B/\xi_b= 
Tf^2\phi^{-3}\ell_Ba^{-1}.
\en  
Here the salt density is assumed not to exceed  the  
density of the charged monomers $n_p$.  
Obviously, the condition $\epsilon_b< T$ yields Eq.(2.1). 
In our case the Debye-H$\ddot{\rm u}$ckel  
 wave number $\kappa \propto  (n_p\ell_B)^{1/2}$ 
is sufficiently small such that 
$n_p\gg \kappa^3$ holds and 
a free energy contribution due 
to the charge density fluctuations 
$\propto \kappa^3$ is negligible 
\cite{Landau,Onukibook}.

\subsection{Ginzburg-Landau free energy} 

For weakly ionized polyelectrolytes, 
we set up the free energy 
accounting  for the 
molecular interactions and the ionization 
equilibrium. We neglect the image 
interaction \cite{Rubinstein,OnukiPRE,Onsager} 
and the  formation  
of  dipole pairs and ion clusters \cite{pair,Krama}. 
The former is important  across an interface 
when the dielectric constants of the two 
phases are distinctly different in the dilute limit 
of the ion densities (see comments in 
the last section), 
while the latter  comes into play  at not small  
ion densities.

The volume fractions   
of the polymer and the solvent 
are written as $\phi({\bi r})$ and 
$1-\phi({\bi r}), 
$ respectively. For simplicity, 
we neglect the volume fractions of the ions 
and  assume that the  monomers  and the solvent molecules have 
a common volume $v_0=a^3$. Then $\phi$ is also the molar 
composition. The counterion  density is written as $n_c({\bi r})$. We 
may add salt with cation and anion densities  
 $n_1({\bi r})$ and $n_2({\bi r})$, respectively.   
The ion  charges are $Z_i$ with $i=c, 1$, and 2. 
In the monovalent case, for example, we have 
$Z_c=1$,  $Z_1=1$,  and $Z_2=-1$, respectively. 
In the continuum limit 
these variables are smooth   coarse-grained ones  
on the microscopic level.

The number of the ionizable monomers on a chain 
is $\nu_M N$ with $\nu_M < 1$. 
In this work the degree of ionization  
  $\zeta({\bi r})$ in the range 
$0\le \zeta \le 1$ depends on  the surrounding 
conditions and is inhomogeneous. Then  the fraction of 
ionized monomers is 
\be 
f=\nu_M \zeta, 
\en 
and the number density of ionized monomers is 
\be 
n_p= v_0^{-1}f \phi= 
v_0^{-1}\nu_M  \zeta \phi,
\en 
where $\zeta$ and $\phi$ are space-dependent. 
The  condition of weak ionization 
$f\ll 1$ is always satisfied for $\nu_M\ll 1$,  
but   we need to require $\zeta\ll 1$  for   $\nu_M \sim 1$. 
Furthermore, we assume that the charge of 
each ionized group is 
negative and monovalent (or equal to $-e$) 
so that the total charge density is written as 
\be 
\rho=e\sum_{i=c, 1, 2}Z_in_i -e n_p .
\en 
The overall charge neutrality condition 
is $\int d{\bi r}\rho=0$. In more detail in 
the present of salt, we have 
\be 
\int d{\bi r}[Z_cn_c  -  n_p]=
 \int d{\bi r}[Z_1n_1  +Z_2  n_2]=0. 
\en  


The free energy  $F$ of our system is the space integral 
of a free energy density $f_T$ in the  fluid container. 
We assume that $f_T $ is  of the form, 
\bea 
f_T&=& f_0(\phi,T) + \frac{T}{2}C(\phi)|\nabla \phi|^2+
\frac{\ve(\phi)}{8\pi}{{\bi E} }^2 \nonumber\\
&& \hspace{-1cm}+ 
T \sum_{i=c, 1, 2} {n_i} [\ln (n_iv_0)-1 +g_i\phi] 
+T (\Delta_0+g_p\phi )n_p  \nonumber\\
&&\hspace{-1cm}
 +Tv_0^{-1}\nu_M\phi[\zeta\ln\zeta+(1-\zeta)\ln(1-\zeta)].
\ena   
In the first  line, the first term $f_0$ is the chemical part in  
the Flory-Huggins form \cite{PG,Onukibook},  
\be 
f_0 = \frac{T}{v_0}[ 
 \frac{\phi}{N}  \ln\phi + (1-\phi)\ln (1-\phi) 
+ \chi \phi (1-\phi) ], 
\en 
where $\chi$ is the interaction parameter 
dependent  on the temperature $T$ 
and its  critical value is $\chi_c= (1+N^{-1/2})^2/2$.  
in the absence of ions. 
The  second term is the gradient part with the composition-dependent 
coefficient \cite{PG},
\be 
C(\phi)=1 /12a \phi(1-\phi),
\en  
where $a=v_0^{1/3}$. The third term is the 
electrostatic free energy, where  ${\bi E}= -\nabla \Phi$ 
is the electric field. 
The electrostatic potential $\Phi$ satisfies   
the Poisson equation,
\be 
\nabla\cdot\ve(\phi)\nabla \Phi=- 
 4\pi \rho. 
\en  
The dielectric constant $\ve(\phi)$ 
changes from the solvent value $\ve_0$ 
to  the polymer value $\ve_p$ with increasing $\phi$. 
For simplicity, we assume 
  the linear form,   
\be 
\ve(\phi)=\ve_0  + \ve_1 \phi ,
\en 
where $\ve_1= \ve_p-\ve_0$. In  some binary 
fluid mixtures, the linear form of $\ve(\phi)$ 
has been measured \cite{Debye-Kleboth}. 
For charged gels, Kramarenko {\it et al.} 
pointed out relevance of strong composition-dependence 
of $\ve(\phi)$  in first-order swelling transition  \cite{gel}.  
In the second line of Eq.(2.7), the   
coupling terms $Tg_i \phi n_i$ ($i=c,1,2,p$)  
arise from the molecular interactions among 
the charged  and  uncharged particles, while   
$T\Delta_0$ 
is the dissociation energy in the dilute limit 
$\phi \to 0$.
In the third  line of Eq.(2.7),  we give the 
entropic free energy 
of dissociation\cite{Joanny,Bu1,Bu2}, where $v_0^{-1}\nu_M\phi$ 
is the density of the ionizable monomers.

{\it Solvation interaction.} 
We have introduced the molecular  interaction terms 
($\propto g_i$), which 
will be simply called the solvation interaction terms. 
In low-molecular-weight binary mixtures 
with salt \cite{Onuki-Kitamura,OnukiPRE},  
such  terms arise from the composition-dependence of 
the solvation chemical potential of ions 
$\mu_i^{\rm sol}$, where $i$ represents the ion species. 
The original Born theory \cite{Born} 
gave  $\mu_i^{\rm sol}=Z_i^2e^2/2\ve(\phi)R_i$, where 
$Z_i e$ is the ion charge and  $R_i$  is called 
the  Born radius. Here  $R_i$  is  
 of order $1{\rm \AA}$ for small 
metallic ions in aqueous solution \cite{Is}. 
For each ion species,  the difference of 
$\mu_i^{\rm sol}$ in the  coexisting two phases 
is the Gibbs transfer free energy 
 typically much larger than $T$ per ion   
 ($\sim  15 T$ for monovalent  ions) 
in electrochemistry \cite{Hung,Osakai}. 
In polymer solutions,  the origin of these terms 
can be  more complex \cite{Hakim,Hakim1,Hydrogen}. 
For example, ions interact with the dipoles of 
  the solvent molecules and those on the chains  
differently,   affecting   
 the hydrogen bonding around the chains.  
Therefore, the solvation chemical potential  
of an charged particle of  the species $i$ 
($i=c,1,2,p$)  arises from the interaction 
 with  the solvent 
molecules and that with the uncharged monomers as 
\be 
\mu_i^{\rm sol}=  \epsilon_i^s (1-\phi)+ 
\epsilon_i^m\phi = 
(\epsilon_i^m-\epsilon_i^s)\phi+\epsilon_i^s, 
\en   
where $\epsilon_i^m$ and $\epsilon_i^s$ are 
the interaction energies. 
The solvation contribution to the 
free energy is given by the space integral of 
the sum, 
\be 
 \sum_i   \mu_i^{\rm sol}n_i
= T\sum_i  g_i \phi n_i + \sum_i  \epsilon_i^s n_i.
\en  
Here we find $g_i$ appearing in Eq.(2.7)  expressed as    
\be 
g_i=(\epsilon_i^m-\epsilon_i^s)/T.
\en   
The  last term on the right hand side 
of Eq.(2.12) contributes to a constant 
(irrelevant) chemical potential for $i=1,2$ 
and to the constant $\Delta_0$ in Eq.(2.7) for $i=c, p$ 
(with the aid of Eq.(2.6) for $i=c$).
As a result, we have attraction for   
$g_i<0$ and repulsion for  $g_i>0$ 
between the ions ($i=c,1,2$) and the polymer chains, 
while we have a composition-dependent 
dissociation constant from $i=c, p$ (see Eq.(2.27)).   
For water-like solvent, 
$g_i>0$ for hydrophilic ions and  
$g_i<0$ for hydrophobic  ions.  
The  Born theory \cite{Born} 
and the data of the Gibbs transfer free energy 
\cite{Hung,Osakai} both suggest 
that $|g_i|$  mostly  much exceeds   unity and 
is even larger for multivalent ions 
such as Ca$^{2+}$.

{\subsection{Equilibrium relations}}
As a typical experimental geometry,  
our fluid system is inserted between 
two  parallel metal plates 
with area $S$ and separation distance $L$(much shorter 
than the lateral dimension  $S^{1/2}$). 
If the surface charge densities 
at the upper and lower 
plates are fixed at  $\pm \sigma_{0}$, 
the electrostatic energy 
$F_e=\int d{\bi r}{\ve(\phi)}{{\bi E} }^2/{8\pi}$ 
is a functional of $\phi$ and $\rho$. The potential 
values at the two plates are 
laterally homogeneous, but are fluctuating quantities 
\cite{OnukiPRE}.

For small variations 
  $\delta\phi$ and $\delta\rho$ superimposed on  
 $\phi$ and $\rho$,   
the incremental change 
of $F_e$ is written  as \cite{OnukiPRE}
\be 
\delta F_e=\int d{\bi r}\bigg[\Phi\delta\rho 
 -\frac{\ve_1{\bi E}^2}{8\pi}\delta\phi \bigg].
\en 
Under the charge neutrality $\int d{\bi r}\rho=0$, 
we first minimize $F$ with respect to 
$\zeta$ (or $n_p$)  and $n_c$ at fixed $n_1$, $n_2$, and $\phi$. 
It is convenient to  
introduce  $G= F+(T\lambda/e) \int d{\bi r}\rho$,  
 where $T\lambda/e$ is the Lagrange multiplier 
independent of space. Using Eq.(2.15) we may calculate 
${\delta G}/{\delta \zeta}$ and ${\delta G}/{\delta n_c}$. 
Setting them equal to zero, we  obtain 
\bea 
&&
\frac{\zeta}{1-\zeta}=\exp[U+ \lambda-\Delta_0-g_p\phi], \\
&& n_c=v_0^{-1}  \exp[- U- \lambda- g_c\phi], 
\ena
where we  introduce  the 
normalized electric potential by 
\be 
U= {e}\Phi/T. 
\en 
Next, homogeneity  of the ion chemical potentials 
$\mu_i= \delta F/\delta n_i$ ($i=1,2)$ yields  
\be 
n_i= n_i^0 \exp[-g_i\phi -Z_i(U+\lambda)], 
\en 
where $n_i^0= v_0^{-1}e^{\mu_i/T}$ are constants. 
Here $U$ and $\lambda$ appear 
 in the combination $U+\lambda$ 
in all the physical quantities. 
If $U+\lambda$ is redefined as $U$, 
$\lambda$ may be set equal to zero 
without loss of generality.

We also require 
homogeneity of $h=\delta F/\delta \phi$, 
where we  fix 
 $n_c, n_1,n_2$ and $n_p$ in  the functional derivative. 
With the aid of Eq.(2.15) some calculations give      
\bea 
\frac{h}{T} &=& \frac{1}{T} f_0'(\phi)+ \frac{C'}{2}|\nabla\phi|^2 
 -\nabla \cdot C\nabla \phi  - 
\frac{\ve_1}{8\pi T}{ E}^2\nonumber\\
&&  + \sum_{i=c, 1, 2, p}  g_in_i 
  + v_0^{-1} \nu_M \ln(1-\zeta),
\ena 
where  $f_0'= {\p f_0}/{\p \phi}$ and  
$C' = \p C/\p \phi$. On the right hand side, the first three terms 
are those in the usual Ginzburg-Landau theory \cite{Onukibook}. 
The last three terms arise from the electrostatic interaction, 
the solvation interaction, and the 
dissociation equilibration,  respectively. 
We may calculate the interface profiles 
and the mesophase profiles from the homogeneity of $h$ 
\cite{OnukiPRE}.

{\it{Surface tension}.} 
In the above procedure, we have minimized the grand potential 
$\Omega=\int d{\bi r}\omega$ under the charge neutrality 
$\int d{\bi r}\rho=0$, where the grand potential density 
is defined by 
\be 
\omega=f_T-h\phi-\mu_1n_1-\mu_2n_2, 
\en 
with $f_T$ being given by Eq.(2.7). Using Eqs. (2.16)-(2.19) 
we may eliminate $\mu_1$ and $\mu_2$ to obtain 
\bea 
\omega&=&f_0 + \frac{TC}{2}|\nabla \phi|^2-h\phi +
\frac{\ve}{8\pi}{{\bi E} }^2-\rho\Phi \nonumber\\
&&- 
T \sum_{i=c, 1, 2} {n_i} +Tv_0^{-1}\nu_M\phi \ln(1-\zeta).  
\ena 
Furthermore, using Eq.(2.20) we may 
calculate the space gradient of $\omega$ as 
\be 
\frac{\p\omega}{\p x_k}= 
\sum_\ell \frac{\p}{\p x_\ell}
\bigg ( {TC}\frac{\p \phi}{\p x_k}
\frac{\p \phi}{\p x_\ell}\bigg)- \frac{\p}{\p x_k}\rho U, 
\en  
where $\p/\p x_k$ and  $\p/\p x_\ell$ are the space derivatives 
with respect to the Cartesian coordinates 
$x,y$, and $ z$. In  the one-dimensional case, where all 
the quantities vary along the $z$ axis, the above equation 
is integrated to give 
\be 
\omega= TC(\phi')^2-\rho \Phi +\omega_\infty,
\en 
where $\phi'=d\phi/dz$ and $\omega_\infty$ is a constant. 
Therefore, around a planar interface separating two bulk phases,   
$\omega(z)$ tends to a common constant 
$\omega_\infty$ as $z\to \pm\infty$. 
From the above relation the surface tension $\gamma=
\int dz[\omega(z)-\omega_\infty]$ is expressed as  
\bea 
\frac{\gamma}{T} 
&=&
 2 \int dz\bigg
[\frac{\hat{f}_0 }{T} - 
\sum_{i=c, 1, 2} {n_i} +\frac{\nu_M\phi}{v_0} 
\ln(1-\zeta)\bigg] \nonumber\\ 
&=& \int dz\bigg
[C(\phi')^2-
\frac{\ve(\phi)}{4\pi T} {\bi E}^2\bigg] , 
\ena 
where $\hat{f_0}= 
f_0 -h\phi-\omega_\infty$ in the first line and 
use is  made of $\int dz\rho \Phi= \int dz \ve {\bi E}^2/4\pi$. 
In the second line the integrand  
consists of a positive gradient term and a  negative 
electrostatic term. Similar expressions 
for the surface tension have been obtained for 
 electrolytes \cite{OnukiPRE} 
and ionic surfactant systems \cite{OnukiEPL}.

{\it{Mass action law}}.
If Eqs.(2.12) and (2.13) are multiplied,  $U$ cancels to disappear. 
It follows   the equation of ionization equilibrium 
or the mass action equation \cite{Bu1,Bu2}, 
\be 
\frac{\zeta}{1-\zeta}n_c =K(\phi),   
\en 
where $K(\phi)$ is the 
 dissociation constant of the form, 
\be 
K(\phi) =v_0^{-1} \exp[-\Delta_0-(g_p+g_c)\phi]. 
\en 
We may interpret   $\Delta(\phi) \equiv 
\Delta_0+ (g_p+g_c)\phi$ 
as  the composition-dependent  dissociation energy divided by $T$.  
  With increasing $\phi$, the dissociation decreases 
for positive $g_p+g_c$ and increases for negative $g_p+g_c$.
If $g_p+g_c \gg 1$, $K(\phi)$ much decreases 
even for a small increase of $\phi$. 
Then $\zeta$  and $n_p$ are related to $n_c$  as 
\be 
\zeta= \frac{v_0n_p}{\nu_M\phi} 
= \frac{ K}{K+n_c}. 
\en 
These relations  hold  in 
equilibrium states,  which may be inhomogeneous. 
In our theory, $n_p=n_p(\phi,U)$ is a function of  the local values 
of $\phi$ and $U$ as well as $n_c$, $n_1$, and $n_2$. 
Here, 
\be 
\frac{\p n_p}{\p U}= (1-\zeta)n_p,
\quad \frac{\p n_p}{\p \phi}= \frac{n_p}{\phi} - 
(1-\zeta)g_pn_p,
\en 
so  $n_p$ increases with increasing $U$ at fixed $\phi$.

{\it Relations in bulk without salt}. 
Furthermore, in a homogeneous bulk phase with 
$n_p=n_c$,   Eq.(2.28) yields  
 the quadratic equation   for $n_c$,
\be 
n_c({n_c+K})=v_0^{-1}\nu_M\phi K,
\en
which is solved to give  
\be 
\zeta= \frac{v_0n_c}{\nu_M\phi} 
= \frac{2}{\sqrt{Q +1}+1} .
\en 
Here it is convenient to introduce 
\be 
Q(\phi) ={4\nu_M\phi}/{v_0 K(\phi)}.
\en 
In particular, we find 
  $\zeta\ll 1$ and $n_c\cong (\nu_M\phi K/v_0)^{1/2}$ 
for $Q \gg 1$, while 
$\zeta \rightarrow 1$   for $Q \ll 1$.
Thus the ion density $n_c$ has been determined for given $\phi$.

{\it Relations in bulk with salt}. 
As another simple situation, 
we may add a salt whose cations 
are of the same species as the counterions. 
The cations and anions are both monovalent. 
Here the counterionsa and the cations from 
the salt are indistinguishable. Thus 
the  sum of  the counterion density 
$n_c$ and the  salt 
cation density $n_1$ is written 
 as $n_c$, while the  salt anion density is 
written as $n_2$.  The  
charge neutrality condition becomes 
$n_c=n_p+n_2$ in the bulk. From  Eq.(2.28) we obtain 
\be 
n_c=v_0^{-1} {\nu_M\phi}K/(K+ n_c)+n_2,
\en 
in the bulk phase. Then $n_c$ increases with increasing $n_2$. 
We treat  $n_2$ as an externally  given constant to obtain  
\bea 
n_c &=&  \frac{1}{2}(n_2-K)
 + \frac{1}{2}\sqrt{(n_2+K)^2+ 4v_0^{-1} \nu_M \phi K}\nonumber\\
&\cong & n_2 + v_0^{-1} \nu_M \phi K/(n_2 +K),
\ena
where the second line holds in the case  $n_2+K \gg 
2(\nu_M \phi K/v_0)^{1/2}=Q^{1/2}K$.  With  increasing $n_2$, 
the ionized monomer density 
$n_p=n_c-n_2$ decreases, while $n_c$ increases.  
At high salt densities, 
where  $n_2$  much exceeds both 
$2(\nu_M \phi K/v_0)^{1/2}$ and $K$, we 
eventually obtain 
\be 
n_c\cong n_2,\quad   
n_p \cong \nu_M \frac{K\phi}{n_2} \ll n_2, \quad 
\zeta \cong \frac{K}{n_2} \ll 1.
\en

\setcounter{equation}{0}
\section{Structure Factor of Composition Fluctuations }

In phase transition theories \cite{Onukibook} 
the order parameter fluctuations 
obey the equilibrium distribution 
$\propto \exp (- F/T)$, where $F$ is the  
Ginzburg-Landau free energy functional. 
In the present problem, the thermal flucuations 
of $\phi$, $n_i$, and $\zeta$ are assumed to 
obey the distribution $\propto \exp (- F/T)$ 
in equilibrium, where 
 $F$ is the space integral of   $f_T$ in Eq.(2.7).   
In  the  Gaussian  approximation of $F$, 
we   consider   small  plane-wave fluctuations 
of $\phi$, $n_i$, and $\zeta$ 
with wave vector $\bi q$ 
in  a one-phase state. It then follows 
the  mean-field  expression for 
the structure factor $S(q)$ 
of the composition fluctuations. 
It is of the form of Eq.(1.1) for a constant  
degree of ionization $\zeta$ in the absence of 
 the solvation interaction. 
Here it will be calculated 
 including the solvation interaction and 
in the annealed case. 
We examine how it depends on the  parameters 
$g_c$, $g_1$, $g_2$, and $g_p$ and how it is modified  by 
the fluctuating ionization.

\subsection{Gaussian approximation}

 From  Eq.(2.7) 
the  fluctuation contributions  to $F$ 
in the  bilinear order  are  written as 
\bea
\frac{\delta F}{T}   &=&
\sum_{\bi q}  \bigg [
\frac{1}{2}({{\bar r}+C q^2})  |\phi_{\small{\bi q}}|^2+
\frac{2\pi }{\ve q^2}|\rho_{\small{\bi q}}|^2 
+\sum_{i=c,1,2}
\frac{|n_{i{\small{\bi q}}}|^2}{2{n_i}}
\nonumber\\
&&\hspace{-1cm} 
+  \sum_{i=c,1,2,p} g_i n_{i {\small{\bi q}}}\phi_{\small{\bi q}}^* 
+  \frac{n_p }{2(1-\zeta)\zeta^2}|\zeta_{\small{\bi q}}|^2
\bigg ]  ,
\ena 
where   $\phi_{\small{\bi q}}$, $\rho_{\small{\bi q}}$, 
 $n_{i{\small{\bi q}}}$  ($i=c, 1,2,p)$, and $\zeta_{\small{\bi q}}$  are  the Fourier components 
of $\phi({\bi r})$, $\rho({\bi r})$, $n_i({\bi r})$, and  $\zeta({\bi r})$,  
respectively. From Eq.(2.4) the fluctuation 
of the charged-monomer density is of the form,  
\be 
n_{p {\small{\bi q}}}=\frac{n_p}{\zeta} 
\zeta_{{\small{\bi q}}}+ \frac{n_p}{\phi} \phi_{ {\small{\bi q}}}.
\en  
In this section $\phi$ and  $n_i$ denote 
 the spatial averages, where  
$n_p=\sum_{i=c,1,2}Z_in_i$ from the 
overall charge neutrality.  
The inhomogeneity in the dielectric constant 
may be neglected for small fluctuations.  
From the Flory-Huggins free energy (2.8) 
the  coefficient $\bar r$ is of the form,      
\be 
 v_0 {\bar r}= \frac{v_0}{T}\frac{\p^2f_0}{\p\phi^2}=
 \frac{1}{N\phi}+ \frac{1}{1-\phi} -2\chi .
\en 
On the right and side of Eq.(3.1),  the first 
term yields the Ornstein-Zernike structure factor 
of the composition without the coupling to the 
charged particles. The second and third terms   are 
well-known for electrolyte systems,   leading to the 
Debye-H$\ddot{\rm u}$ckel screening of the charge density 
correlation. The fourth term arises from 
the solvation interaction, while  the fifth term  
from the fluctuation of ionization.

 We minimize  $\delta F$ 
with respect to $n_{i {\small{\bi q}}}$   ($i=c, 1,2,p)$ 
at fixed $\phi_{{\small{\bi q}}}$ to express 
them in the linear form $n_{i {\small{\bi q}}}\propto 
\phi_{{\small{\bi q}}}$.  In particular, 
$\rho_{{\small{\bi q}}}$ and 
$\zeta_{{\small{\bi q}}}$ are written as    
\bea 
\rho_{{\small{\bi q}}}&=& \frac{q^2}{4\pi \ell_B}
U_{{\small{\bi q}}}= 
-\frac{Aq^2}{q^2+\kappa_T^2}\phi_{{\small{\bi q}}},\\ 
\zeta_{{\small{\bi q}}}&=& -\zeta(1-\zeta)\bigg [g_p+ 
\frac{4\pi \ell_BA}{q^2+\kappa_T^2}\bigg]\phi_{{\small{\bi q}}}, 
\ena 
where $U_{{\small{\bi q}}}$ is the Fourier 
component  of $U({\bi r})$,   and 
\bea	 
&&\kappa_T^2= 4\pi \ell_B [(1-\zeta)n_p+\sum_{i=c,1,2} Z_i^2n_i],\\
&&A= \frac{n_p}{\phi}  - (1-\zeta)g_pn_p+\sum_{i=c,1,2}Z_ig_i n_i .
\ena 
Here   $\kappa_T$ defined in Eq.(3.6) is a generalized 
 Debye-H$\ddot{\rm u}$ckel  
 wave number introduced by 
Raphael and Joanny \cite{Joanny}, 
where the term proportional to $n_p$   
  arises from  dissociation and recombination 
on the polymer chains.   The first relation (3.4) itself 
readily follows from the relations, 
\be 
\frac{\p \rho}{\p \phi}=-eA,\quad 
\frac{\p \rho}{\p U}=-\frac{e\kappa_T^2}{4\pi \ell_B},
\en 
where $\rho({\bi r})$ is regarded 
as a function of the local values of 
$U({\bi r})$ and $\phi({\bi r})$ as discussed  
around  Eqs.(2.28) and (2.29).  From Eq.(3.5) 
we find $\zeta_{{\small{\bi q}}}\rightarrow 0$ as 
$\zeta \to 0$ or 1. We can see that 
  the last term on the right hand side of  Eq.(3.1) 
is negligible as $\zeta \to 1$.

After elimination of 
 $n_{i {\small{\bi q}}}$   ($i=c, 1,2,p)$  
the  free energy change 
$\delta F$ is expressed as $\delta F 
= \sum_{\bi q}T |\phi_{\small{\bi q}}|^2/S(q)
$, where $S(q)$ is the composition structure factor calculated as 
\bea	
\frac{1}{S(q)}
&=&{\bar r}+ r_s + C q^2 
+ \frac{4\pi \ell_B A^2}{q^2+\kappa_T^{2} } \nonumber\\ 
&&\hspace{-1.5cm}= 
{\bar r}+ r_s+ C\gamma_p^2\kappa_T^2  
+ C q^2\bigg[ 1- 
\frac{\gamma_p^2\kappa_T^2}{q^2+\kappa_T^{2} }\bigg],
\ena
where $r_s$ is the shift of $\bar r$ 
arising  from the solvation 
interaction given by  
\be	 
r_s = [{2}{\phi}^{-1}g_p  -(1-\zeta)g_p^2]n_p- \sum_{i=c,1,2}g_i^2n_i.
\en 
As $\phi\to 0$ we have $r_s \cong 2g_p f$. 
For large $|g_i|\gg 1$ the other negative terms ($\propto g_1^2$) 
can be significant, leading to  $r_s<0$.  
In the second line of Eq.(3.9)  $\gamma_p$ is defined by 
\be 
\gamma_p=
\sqrt{\frac{4\pi\ell_B}{C}}\frac{A}{\kappa_T^2}=  
\frac{(4\pi\ell_BC)^{-1/2} A}{(1-\zeta)n_p+\sum_{i=c,1,2} Z_i^2n_i}, 
\en 
where $(4\pi\ell_BC)^{-1/2}= 
[3a\phi(1-\phi)/\pi\ell_B]^{1/2}$ if use is made of Eq.(2.9). 
Note that $r_s$ and $\kappa_T^2$ 
consist of the terms  proportional 
to the charge densities, while $\gamma_p$ depends on their  ratios. 
In particular, if  $n_c=n_p$ and $n_1=n_2$ in the monovalent case,  
$\gamma_p$ depends on the density ratio $R\equiv n_1/n_p$ as
\be 
\gamma_p =
\frac{(4\pi\ell_BC)^{-1/2}}{{2-\zeta
+2R}} \bigg[\frac{1}{\phi}-(1-\zeta)g_p+g_c+(g_1-g_2)R\bigg].
\en 
Here   $R=0$ for salt-free polyelectrolytes   
and $R$ is increased with increasing the salt density. 
As  $R\to \infty$ the above formula tends to that  
for neutral polymer solutions (low-molecular-weight 
binary mixtures for $N=1$) with salt. 
We note that $\gamma_p$ can even be negative 
depending on the solvation terms in the brackets 
on the right hand side of Eq.(3.12).

\subsection{Macrophase and microphase separation}

From the second line of  Eq.(3.9) 
we obtain the small-$q$ expansion 
$S(q)^{-1}= a_0+ a_1 q^2+ a_2q^4\cdots$ 
with 
\be 
a_1=C(1-\gamma_p^2),
\en
so we encounter  the following two cases. 
(i) If $|\gamma_{\rm p}|<1$, 
$S(q)$ is maximum at $q=0$ and we predict the usual 
phase transition with increasing $\chi$ 
in the mean field theory. The spinodal  
$\chi=\chi_{\rm sp}$ is given   by 
\be 
2\chi_{\rm sp}
= \frac{1}{N\phi}+ \frac{1}{1-\phi}+r_s+ C\gamma_p^2\kappa_T^2.
\en 
Macroscopic phase separation occurs 
for $\chi>\chi_{\rm sp}$. 
On  the right hand side of Eq.(3.14),  
$r_s=0$ and the last term becomes $n_p/(1-\zeta)\phi^2>0$ 
for $g_i=0$, but the sum of 
the last two terms can be  much altered 
for  $|g_i|\phi \gs 1$. For example, without salt and 
in the monovalent case, it is equal to 
$[1/\phi^{2}+\zeta g_c^2+2g_c/\phi]n_p/(1-\zeta) 
-2g_pg_cn_p$. 
(ii)  If $|\gamma_{\rm p}|>1$,  
  $S(q)$ has a peak at an intermediate wave number 
$q_m$ given by 
\bea 
q_m^2&=& (4\pi \ell_B/C)^{1/2}|A| - \kappa_T^2\nonumber\\
&=&\kappa_T^2(|\gamma_p| -1) . 
\ena  
Thus $q_m \to 0 $ as $\kappa_T \to 0$ or as $|\gamma_p| \to 1$. 
The spinodal  is given by $S(q_m)^{-1}=0$ or by 
\be 
2\chi_{\rm sp}
= \frac{1}{N\phi}+ \frac{1}{1-\phi}+r_s
+C\kappa_T^2 (2|\gamma_p|-1). 
\en  
Microphase separation should be triggered  
for $\chi>\chi_{\rm sp}$. 
Since $\gamma_p$ depends only on the ion density 
ratio $R$ as in Eq.(3.12), 
the criterion $\gamma_p>1$ 
remains unchanged however  small the densities of the charged 
particles are.  
Of course, mesophase formation is 
well-defined only when $q_m$ in Eq.(3.15) is much larger than 
the inverse of the system length. 
In the following we discuss some special cases.

{\it Polyelectrolytes  without  solvation interaction}. 
For $g_i=0$,  we have $r_s=0$ and  
$A=n_p/\phi= v_0^{-1}f$ as in the previous 
theories of weakly ionized 
polyelectrolytes \cite{Lu1,Lu2,Joanny}. 
In the monovalent case  Eq.(3.12) gives  
\bea 
\gamma_p &=& 1/[(4\pi\ell_BC)^{1/2}{(2-\zeta+2R)\phi}]\nonumber\\
 &=& [3a \phi /\pi\ell_B(1-\phi)]^{1/2}/(2-\zeta+2R), 
\ena 
where $R=n_1/n_p$ and use is made of Eq.(2.9) in the second line.  
Here $\gamma_p$ is largest without salt 
($R=0$)  and decreases with increasing $R$. 
In accord with this predicted salt effect,  
Braun et al. \cite{Candau} observed 
a mesophase    at low  salt concentration and 
a  macroscopic 
phase  separation at high salt concentration. 
However, for large $|g_i|$, the solvation interaction  
comes into play. 
Thus more experiments are desirable to detect it 
in  polyelectrolytes with water-like solvent.

{\it Binary mixtures with solvation interaction}. 
 Without ionization or for $f=0$,  we have $n_p=n_c=0$. 
Here we describe  a neutral binary mixture with 
  salt. Further for a  monovalent salt,  we have 
 $A= (g_1-g_2)n_1$ and 
\bea 
\gamma_p &=&  (g_1-g_2)/(16\pi\ell_BC)^{1/2},\\
r_s &=& -(g_1^2+g_2^2)n_1.   
\ena  
From  neutron scattering,  
Sadakane {\it et al.}\cite{Seto}
  found  periodic structures  
in a near-critical,  low-molecular-weight 
 mixture of D$_2$O and trimethylpyridine (3MP)  
containing sodium tetrarphenylborate (NaBPh$_4$). 
Their  salt is composed of  
strongly hydrophilic cation Na$^+$   
and strongly hydrophobic anion 
BPh${_4}^{-}$.  If their data are interpreted 
in our  theoretical scheme,  
we expect $g_1 \gg 1$, $g_2 \ll  -1$, 
and $\gamma_p>1$ for their system. 
Such ion pairs are antagonistic to each other, undergoing 
microphase separation where the solvent composition  
is inhomogeneous.  
It is worth noting that  the scattering amplitude 
was maximum at  $q=0$ upon addition  of 
NaCl,  KCl, etc. in the same 
mixture D$_2$O-3MP \cite{Seto1}. 
These  salts consist of hydrophilic anions and cations 
with $g_1 \sim g_2$, so $\gamma_p$ 
should be  much smaller leading to no 
mesophase formation. 
It is also striking that 
the coexistence curve is much shifted 
as  $\Delta T \propto r_s 
\propto g_i^2$ for large $|g_i|$ 
 with increasing the salt density 
even for hydrophilic ion pairs,  
which is  consistent  with a number of previous 
experiments \cite{polar1,polar2,newly_found}.

{\it Neutral polymer solutions with salt}. 
For very weak ionization but in the presence of salt,   
$\gamma_p$ is expressed as in Eq.(3.18). 
In our theory, even   polymer solutions   
consisting of neutral polymers 
and a polar solvent can 
exhibit a charge-density wave  phase 
for  $|g_1-g_2|\gg 1$. 
Hakim {\it et al}. \cite{Hakim,Hakim1} found  
a broad  peak  at an intermediate 
wave number in the scattering amplitude 
in (neutral)  polyethylene-oxide (PEO) 
solutions with methanol and with acetonitrile 
by adding a small amount of  salt KI. 
They ascribed the origin of the peak to 
binding of potassium ions  to PEO chains. 
In our theory such a peak can arise 
for a  sufficiently large  $g_1-g_2$.  
Remarkably, the peak 
 disappeared if the solvent was water, 
which indicates  sensitive 
 dependence of the molecular interactions 
on  the solvent species.  Thus more experiments 
should be  informative, where 
use of antagonistic ion pairs will yield   
more drastic effects, leading to mesophases.

\setcounter{equation}{0}
\section{One-dimensional profiles and numerical results}

We give  numerical results 
of one-dimensional profiles in equilibrium. 
All the quantities are assumed to vary along the  
$z$ axis. For each given  $\phi(z)$, 
we need to  solve the nonlinear Poisson equation 
(2.10) numerically. The charge density is expressed as
\be 
\rho(z) =e \sum_{i=c,1,2}Z_i n_i(z)-
\frac{e \nu_M\phi(z)}{1+n_c(z)/K(z)},
\en 
where  $n_i(z)\propto 
\exp[-g_i\phi(z)-Z_i U(z)]$ 
from Eqs.(2.17) and (2.19) and $K(z)$ depends on $\phi(z)$ as 
in Eq.(2.27). 
We then seek $\phi(z)$ self-consistently 
from the homogeneity of the right hand side of Eq.(2.20).

As in electrolytes, there appears a difference 
of the electric potential $\Delta \Phi$ 
across an interface, which is called the 
Galvani potential difference \cite{OnukiPRE}. 
Because there are many parameters, 
we will set $\chi=1$, $N=20$,   
$\ve_1=-0.9\ve_0$, and $\ell_B=e^2/\ve_0T= 
8a/\pi$ in 
all the following examples.   Then the  dielectric constant 
of the solvent is 10 times larger than 
that of the polymer. The space will
 be measured in units of the 
molecular size $a=v_0^{1/3}$.

\subsection{Interface  without salt}

\begin{figure}[h]
 \includegraphics[scale=0.5]{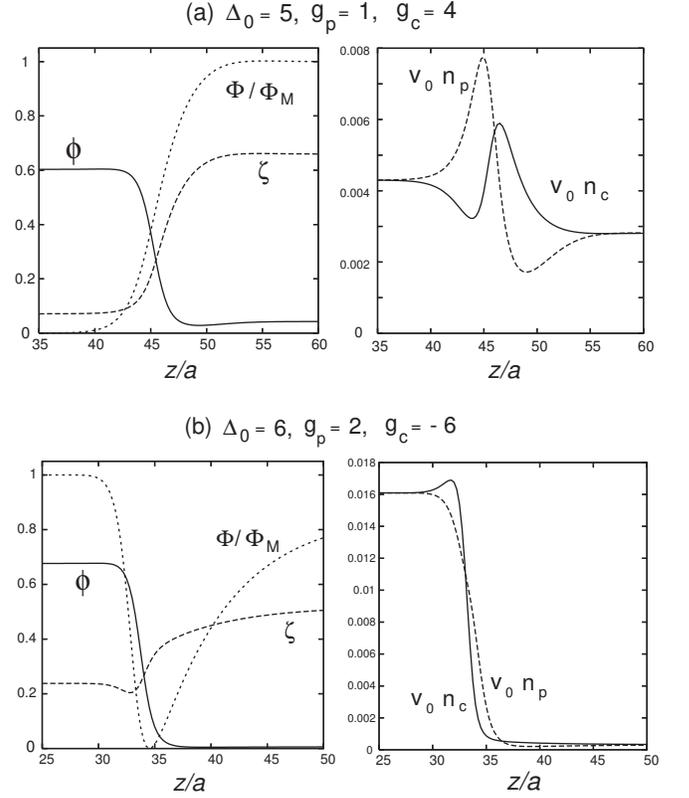} 
\caption{Interface profiles in the salt-free case 
for  (a) $\Delta_0=5$, $g_p=1$, and $g_c=4$ (top)  
and for (b) $\Delta_0=8$, $g_p=2$, and $g_c=-6$ (bottom). 
Polymer volume fraction $\phi(z)$, 
normalized potential $\Phi(z)/\Phi_M$, 
and degree of ionization $\zeta(z)$ (left), and 
normalized charge densities $v_0n_c(z)$ and  $v_0n_p(z)$ 
(right). The other parameters 
are common as  $\chi=1$,  $N=20$, $\nu_M=0.1$, 
$\ve_1=-0.9\ve_0$, and $\ell_B=8a/\pi$. 
Here $\Phi(z)$ is measured from its  
minimum,  and $\Phi_M (=2.67T/e$ in (a) and $6.86T/e$ in (b)) is 
the difference of its maximum and  minimum. 
}
\end{figure}

First  we suppose coexistence of 
 two salt-free phases,  separated by a planar 
interface. Hereafter the quantities with the subscript $\alpha$ 
(the subscript $\beta$) denote 
the bulk values in the polymer-rich (solvent-rich)  phase  
attained as  $z \rightarrow -\infty$ 
(as  $z \rightarrow \infty$). Namely, 
we write $n_{c\alpha}= 
n_c(-\infty)$ and $n_{c\beta}= n_c(\infty)$. 
Here we obtain $n_{c\alpha}$ or  $n_{c\beta}$ from 
 Eq.(2.29) and  
\be 
\frac{n_{c\alpha}}{
n_{c\beta}}= \frac{\phi_\alpha\zeta_\alpha}{\phi_\beta\zeta_\beta}=
\exp({-g_c\Delta\phi-\Delta U})
\en 
from Eq.(2.17). 
Hereafter  $\Delta\phi=\phi_\alpha-\phi_\beta$ is the 
composition difference and 
\be 
\Delta U=e\Delta\Phi/T 
=e(\Phi_\alpha-\Phi_\beta)/T
\en  
is the normalized Galvani potential difference. 
In terms of $Q_\alpha=4\nu_M\phi_\alpha/v_0K_\alpha$ 
and  $Q_\beta=4\nu_M\phi_\beta/v_0K_\beta$,  we obtain   
\bea 
\Delta U &=& g_p\Delta\phi+\ln \bigg[
\frac{\sqrt{Q_\beta+1}-1}{\sqrt{Q_\alpha+1}-1}\bigg],\\
\frac{n_{c\beta}}{n_{c\alpha}}
&=&\frac{\phi_\beta(\sqrt{Q_\alpha+1}+1)}{\phi_\alpha(\sqrt{Q_\beta+1}+1)}.
\ena 
In particular, if  
$Q_\alpha\gg 1$ and $Q_\beta\gg 1$ 
(or $\zeta_\alpha\ll 1$ and $\zeta_\beta\ll 1$), we obtain 
$
\Delta U \cong (g_p-g_c)\Delta\phi/2+\ln(\phi_\beta/\phi_\alpha)$ 
and 
$
{n_{c\beta}/}{n_{c\alpha}}\cong  
\exp[(g_p+g_c)\Delta\phi/2](\phi_\beta/\phi_\alpha)^{1/2}$.

The interface profiles are  
extremely varied,  sensitively depending   
on the molecular interaction parameters, 
$\Delta_0$, $g_p$, and $g_c$. As such examples, 
in Fig.1, we show salt-free interface profiles  for 
(a) $\Delta_0=5$, $g_p=1$, and $g_c=4$   
and (b) $\Delta_0=8$, $g_p=2$, and $g_c=-6$.  
 In  the $\alpha$ and  $\beta$ regions, 
 the degree of ionization $\zeta$ 
is $0.071$ and $0.65$ in (a) 
and is $0.24$ and $0.51$ in (b), respectively. 
The charge densities 
are multiplied by $v_0=a^3$.   
The $\Delta U=e(\Phi_\alpha-\Phi_\beta)/T$ 
is $-2.67$ in (a) and $0.099$ in (b). 
Interestingly, in (b), 
$U(z)$ exhibits a deep minimum with 
$U_{\rm min}=U_\beta -5.98$ at the interface position.  
We can see appearance of the charge density 
$n_c-n_p$ around the interface, resulting in an electric 
double layer. The counterion density is 
shifted to the $\beta$ region in (a) 
because of positive $g_c$ 
and  to the $\alpha$ region in (b) 
because of negative $g_c$.  
The parameter $\gamma_p$ 
in Eq.(3.11) is $0.75$ in (a) and $0.20$ in (b) 
in the $\alpha$ region, ensuring the stability 
of the $\alpha$ region.  
From Eq.(2.25) the surface tension $\gamma$ is calculated as 
$\gamma=0.0175T/a^2$ in (a) and as $0.0556T/a^2$ in (b), 
while we obtain    $\gamma= 0.050T/a^2$ 
without ions at the same $\chi=1$. 
 In (a) the surface tension $\gamma$ 
is largely decreased because 
the electrostatic term in Eq.(2.25) 
is increased due to the formation of a large 
electric double layer. In (b), on the contrary,  it is increased by 
$10\%$ due to depletion of the charged particles  
from the interface \cite{OnukiPRE}.

\subsection{Interface  with  salt}

\begin{figure}[t]
 \includegraphics[scale=0.41]{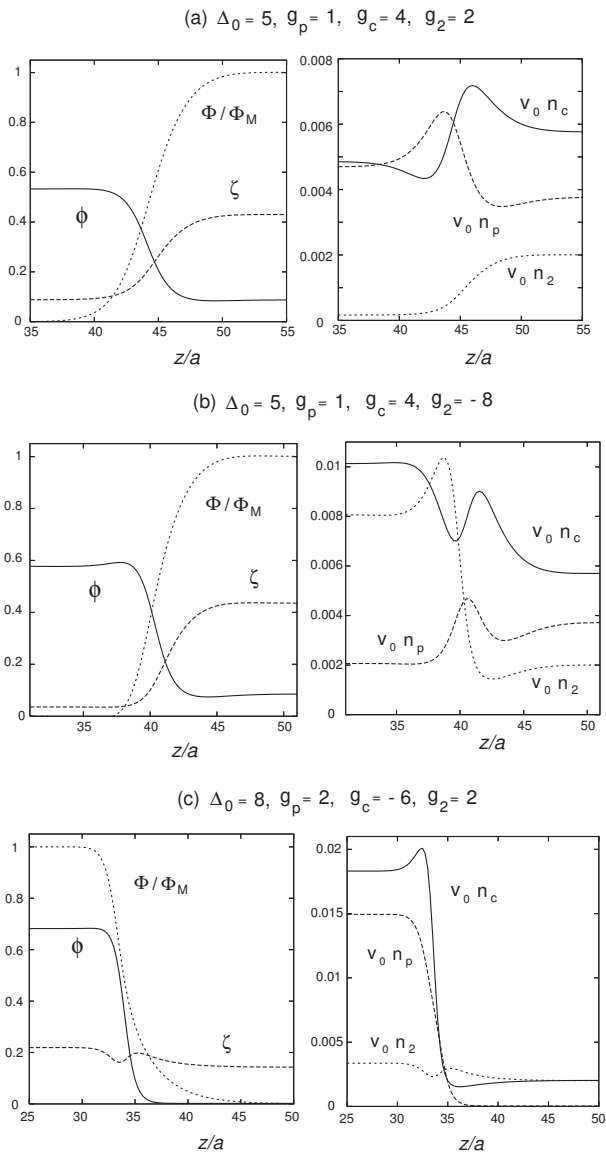} 
\caption{
Interface profiles with mobile cations and  anions 
for  (a) $\Delta_0=5$, $g_p=1$, $g_c=4$, and $g_2=2$ (top),   
for (b) $\Delta_0=5$, $g_p=1$, $g_c=4$, and $g_2=-8$ (middle),   
and for (c) $\Delta_0=8$, $g_p=2$,  $g_c=-6$, and $g_2=2$ (bottom).  
Curves represent   $\phi(z)$, 
 $\Phi(z)/\Phi_M$, 
and  $\zeta(z)$ (left)   and 
 $v_0c_c(z)$,  $v_0c_p(z)$, and  $v_0c_2(z)$ 
(right) with $v_0n_{2\beta}=0.002$. 
Here $\Phi(z)$ is measured from its  
minimum,  and $\Phi_M$ is 
the difference of its maximum and  minimum. 
The other parameters are the same as those in Fig.1. 
}
\end{figure}
  
With addition of salt,  
interface profiles are even more complex. 
For simplicity,  we consider 
a salt whose  cations are 
of the same species as the counterions from the polymer. 
This is the example  discussed at the end of Section 2. 
The free energy density in this case 
is still given by Eq.(2.7) if we  set $n_1=0$ there. 
The densities of the mobile cations, the mobile anions, 
and the charged monomers are written as $n_c(z)$, $n_2(z)$, 
and $n_p(z)$, respectively. 
All the charged particles are monovalent. 
We treat $n_{2\beta}$ in the solvent-rich $\beta$ region 
as a control parameter, which is the salt density added 
in the $\beta$ region. 
The densities  $n_2(z)$ and $n_c(z)$ depend on $z$ as   
\bea 
n_2(z)&=& n_{2\beta}\exp[g_2(\phi_\beta-\phi(z)) 
-U_\beta+U(z)], \nonumber\\
n_c(z)&=& n_{c\beta}\exp[g_c(\phi_\beta-\phi(z)) 
+U_\beta-U(z)].
\ena   
We write $S\equiv n_{2\beta}/K_\beta$ to avoid the cumbersome notation in 
the following. Use of the first line of Eq.(2.34) gives 
\be 
n_{c\beta}=\frac{K_\beta}{2}
\bigg[S-1+ \sqrt{(S+1)^2+Q_\beta}\bigg].
\en  
We  use Eq.(2.33) in the $\alpha$ region 
to determine $n_{c\alpha}$. 
Since $n_{c\alpha}n_{2\alpha}/K_\alpha
=n_{c\beta}n_{2\beta}/K_\beta$ from Eq.(4.5), $n_{c\alpha}$ obeys    
\be 
(n_{c\alpha} -Sn_{c\beta}{K_\alpha}/{n_{c\alpha}})
({1+ n_{c\alpha}/K_\alpha})= {v_0^{-1}\nu_M\phi_\alpha}.
\en  
In particular, if $n_{c\alpha}\gg K_\alpha$ 
or if $\zeta_\alpha\ll 1$, 
we find 
\bea 
&&\hspace{-1cm}n_{c\alpha}
\cong K_\alpha^{1/2}(Sn_{c\beta}+v_0^{-1}\nu_M\phi_\alpha )^{1/2}, \\
&&\hspace{-1cm}\Delta U\cong g_p{\Delta\phi}+
\frac{1}{2}\ln\bigg
[\frac{n_{c\beta}^2/K_\beta^2}{Sn_{c\beta}/K_\alpha+
Q_\alpha}\bigg].
\ena

In Fig.2, we show  profiles 
of $\phi(z)$, $\Phi(z)$, $\zeta(z)$,  $n_c(z)$, $n_2(z)$, 
and $n_p(z)$ with $n_{2\beta}=0.002v_0^{-1}$ held fixed  
for three cases: 
(a) $\Delta_0=5$, $g_p=1$, $g_c=4$, and $g_2=2$;    
(b) $\Delta_0=5$, $g_p=1$, $g_c=4$, and $g_2=-8$;   
(c) $\Delta_0=8$, $g_p=2$,  $g_c=-6$, and $g_2=2$.  
The normalized potential difference $\Delta U$ is 
 $-1.61$ in (a), $ -2.54$ in (b), and 
$1.88$ in (c), while the normalized surface tension $a^2\gamma/T$ 
is given by $0.0358$ in (a), $0.0233$ in (b), and 
$0.0554$ in (c). 
In (a) the cations and anions are both 
repelled from the interface into the $\beta$ region 
because of positive $g_c$ and $g_2$. 
In (b), owing to  $g_2=-8$, 
 the anions become richer 
 in the $\alpha$ region than  in the $\beta$ region,  
since  they are strongly attracted to the 
polymer chains.  There, despite $g_c=4$, the cations 
become  also richer in the $\alpha$ region  to satisfy the 
charge neutrality.   
In (c), we set  $g_c=-6$,  so the cations are 
strongly attracted to the polymer chains. 
The cations are also attracted into the $\alpha$ region. 
The  electric dipole due to the 
double layer is reversed, yielding $\Delta  U= 
U_\alpha-U_\beta>0$.

\subsection{ Periodic state} 
\begin{figure}[h]
 \includegraphics[scale=0.4]{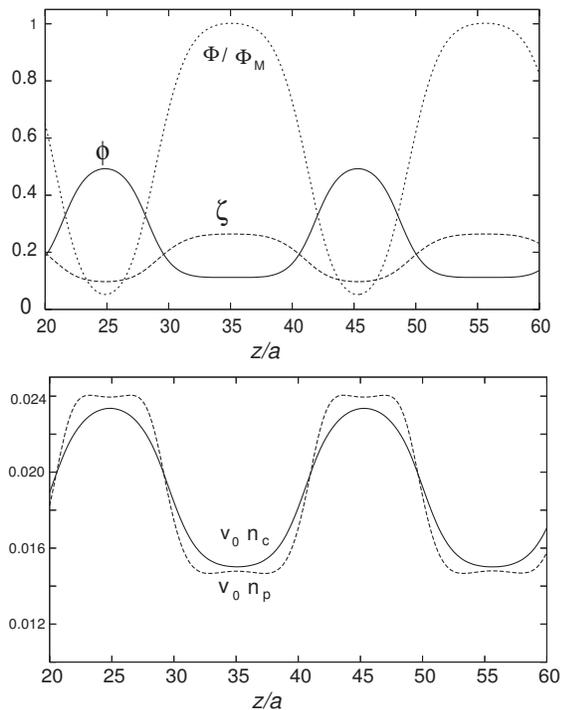} 
\caption{Periodic profiles in a salt-free 
mesophase  with   $\nu_M=0.5$,   $\Delta_0=5$, 
$g_p=1$, and $g_c=4$. 
  Top: $\phi(z)$. 
$\Phi(z)/\Phi_M$ with $\Phi_M=0.86T/e$, 
and  $\zeta(z)$.  Bottom: 
$v_0n_c(z)$ and  $v_0n_p(z)$. 
}
\end{figure}

With varying the temperature (or $\chi$), 
the average composition $\av{\phi}$, 
the amount of salt, 
there can emerge a number of mesophases 
sensitively depending on the various 
molecular parameters ($g_i$, $\Delta_0$, and $\nu_M$). 
In Fig.3, we show an example of a one-dimensional 
periodic state without salt. Here   $\nu_M$ is 
set equal to  $0.5$ and the charge 
densities are much  increased.  In this case, 
the  degree of segregation 
and the charge heterogeneities 
are much   milder than in the cases in  Fig.1. 
Since  the charge density $\rho(z)$ 
 everywhere remains small,  
 Eq.(2.31) locally holds approximately. Thus, 
\be
n_c(z) 
\cong \frac{\nu_M\phi(z)}{\sqrt{Q(z)+1/4}+1/2},
\en
where $Q(z)$ is defined  by Eq.(2.32) and $n_p(z)$ 
is determined by  Eq.(2.26). 
The deviation of $n_c(z)$ from the right hand side 
of Eq.(4.10) gives $\Phi(z)$ as a first approximation. 
The above relation   approximately 
holds slightly below the transition 
from a disordered state 
to a charge-density-wave state.

\section{Summary and remarks}

The charge distributions  
in polyelectrolytes are extremely complex 
around interfaces and in mesoscopic states, 
sensitively depending on the 
molecular (solvation) interaction   
and the dissociation process. 
Our continuum theory takes  account 
of these effects  in the  simplest  manner, 
though it  should be inaccurate 
on the angstrom scale.

Our main results are as follows. 
In Section 2,  we have presented 
a continuum theory accounting for 
the solvation effect and the dissociation 
equilibrium. The degree of ionization is rather a 
dynamic variable dependent on the local electric potential.  
The surface tension in Eq.(2.25) consists of 
a positive gradient term and a  negative electrostatic 
term, where the latter is significant 
as the electric double layer is enhanced as in 
the top panels in Figs.1 and 2.  
In Section 3, we have calculated the structure factor $S(q)$  
of the composition fluctuations for polymer solutions 
with the solvation and ionization effects included. 
When the  parameter $\gamma_{\rm p}$ 
 in Eq.(3.11) or (3.12) exceeds unity in a one-phase state, 
 a peak  can appear at an intermediate wave number 
in  $S(q)$. In such systems, 
a mesoscopic phase can emerge as the interaction 
parameter $\chi$ is increased. 
In Section 4, we have numerically solved the equilibrium 
equations in Section 2 to examine one-dimensional 
profiles of interfaces with and without salt 
and in a periodic state without salt. 
Though very preliminary and fragmentary, 
our numerical results  demonstrate 
dramatic influence  of the solvation interaction 
on inhomogeneous structures.

A number of 
complex effects should be further 
taken into account to properly describe 
polyelectrolytes. 
(i) Under Eq.(2.1) we have neglected 
  the effect of the 
electrostatic interaction on the chain conformations 
to use  the Flory-Huggins free energy density. 
Hence, as argued in the literature \cite{Barrat,Rubinstein}, 
our theory cannot be justified at small $\phi$. 
(ii) We have neglected 
the image interaction, which  is known to increase 
the surface tension of a water-air interface 
at low ion densities \cite{Onsager}. 
Generally, it arises from 
a deformation of the  self-interaction  
of ions  due to  inhomogeneous  $\ve$ \cite{OnukiPRE}. 
So it originates from  the discrete nature of ions and 
is not accounted for in the free energy (2.7), where 
 the electric field $\Phi$ is produced by the smoothly 
 coarse-grained charge density $\rho$. 
(iii) At sufficiently low ion 
densities, we have neglected the ionic correlations 
giving rise to ion dipoles and clusters \cite{pair,Krama}  
and a free energy contribution due 
to the charge density fluctuations 
($\propto \kappa^3$) 
\cite{Landau,Onukibook}. 
(iv) The phase diagram of mesophases 
including the solvation interaction remains 
unknown.  To describe the mesophases 
we need to perform nonlinear analysis 
and computer simulations.  
(v) We have assumed a one-component 
water-like solvent. For two-component solvents 
we may predict a variety of effects. 
For example, 
with addition a small amount of  water  
in a less polar solvent, hydration  shells 
are formed around the hydrophilic 
ionized monomers and the 
counterions \cite{Is,Osakai}, leading to an   increase of 
the degree of ionization. 
For not small water concentrations, 
we may well predict  formation 
of complex mesoscopic structures mediated by 
the Coulombic and  solvation interactions.

\acknowledgments
{
This work was supported in part by Grant-in-Aid for Scientific Research
on the Priority Area "Soft Matter Physics" from the MEXT of Japan. 
}



\begin{references}
\bibitem{PG} P.G. de Gennes,  
{\it Scaling Concepts in Polymer Physics} 
(Ithaca,  Cornell Univ. Press) 1980. 


\bibitem{Barrat} J.L. Barrat and J.F. Joanny, 
 Adv. Chem. Phys. XCIV, I. Prigogine, S.A. Rice Eds.,
John Wiley $\&$ Sons, New York 1996.  


\bibitem{Rubinstein} A.V. Dobrynin and M. Rubinstein, 
 Prog. Polym. Sci. {\bf 30}, 1049 (2005).  

\bibitem{Yoshikawa}  E.Yu Kramarenko, A.R. Khoklov and K. Yoshikawa, 
Macromolecules {\bf 30}, 3383 (1997). 


\bibitem{Joanny} E. Raphael and 
 J. F. Joanny, Europhys. Lett.  {\bf 13}, 623 (1990).

\bibitem{Bu1} I. Borukhov, D. Andelman, and 
H. Orland, Europhys. Lett.{\bf 32}, 499 (1995). 


\bibitem{Bu2}  I. Borukhov, D. Andelman, R. Borrega, M. Cloitre, 
L. Leibler,  and 
H. Orland, J. Phys. Chem. B  
{\bf 104}, 11027 (2000). 



\bibitem{Is} J. N. Israelachvili,  
{\it Intermolecular and Surface 
Forces} (Academic Press, London, 1991). 


\bibitem{Lu1} V. Yu. Borye and I. Ya. Erukhimovich, 
Macromolecules {\bf{21}}, 3240 (1988). 


\bibitem{Lu2} 
 J. F. Joanny and L. Leibler, 
J. Phys. (France) {\bf 51}, 547 (1990).




\bibitem{PD} D. E. Domidontova, I. Ya.  Erukhimovich, 
and A. R. Khokhlov,  
Macromol. Theory Simul. {\bf 3}, 661 (1994). 



\bibitem{Krama} 
E. Yu.  Kramarenko, I. Ya.  Erukhimovich, 
and A. R. Khokhlov,  
Macromol. Theory Simul. {\bf 11}, 462 (2002). 




\bibitem{Onuki-Kitamura} A. Onuki and H. Kitamura, 
  J. Chem. Phys. {\bf 121}, 3143 (2004).

\bibitem{OnukiPRE} A. Onuki, Phys. Rev. E {\bf 73}, 021506 (2006);  
J. Chem. Phys.  {\bf 128}, 224704 (2008).
\bibitem{OnukiEPL} A. Onuki, 
EPL, {\bf 82}, 58002 (2008). 
\bibitem{OnukiW} A. Onuki, 
to be published in: "Polymers, 
Liquids and Colloids in Electric Fields: Interfacial Instabilities, Orientation, and Phase-Transitions", Eds. Y. Tsori and U. Steiner, World Scientific (2008). 


\bibitem{Nishida} K. Nishida, K.  Kaji, and T.  Kanaya, 
Macromolecules {\bf 28}, 2472 (1995). 

\bibitem{Shibayama} 
M. Shibayama, T. Tanaka, J. Chem. Phys. {\bf 102}, 9392
(1995).





\bibitem{Candau} O. Braun, F. Boue, and F. Candau, 
Eur. Phys. J. E {\bf 7}, 141 (2002).  

\bibitem{Seto} K. Sadakane, H. Seto, H. Endo, and M. Shibayama, 
J. Phys. Soc. Jpn., {\bf 76}, 113602 (2007). 


\bibitem{Hakim} I.F. Hakim and J. Lal, Europhys. Lett. 
{\bf 64},  204  (2003).

\bibitem{Hakim1} I.F. Hakim, J. Lal, and M. Bockstaller, 
Macromolecules {\bf 37},  8431 (2004).





\bibitem{Shi} Shi, A.-C.; Noolandi, J. 
Maromol. Theory Simul. {\bf 1999}, 8(3), 214. 


\bibitem{Taniguchi} 
Q. Wang, T. Taniguchi, and G.H. Fredrickson, 
J. Phys. Chem B {\bf{2004}}, 108, 6733-6744; 
ibid.  {\bf{2005}}, 109, 9855-9856.


\bibitem{Landau} 
L.D. Landau and E.M. Lifshitz, {\it Statistical Physics} (Pergamon, New York, 1964). 

\bibitem{Onukibook} A. Onuki, {\it Phase Transition Dynamics} 
(Cambridge University Press, Cambridge, 2002).



\bibitem{Onsager} 
L. Onsager and N. N. T. Samaras, J. Chem. Phys. {\bf 2}, 528 (1934).




\bibitem{pair} L. Degr$\rm\grave{e}$ve and F.M. Mazz$\rm{\acute{e}}$, 
Molecular Phys. 
{\bf 101}, 1443 (2003);  A.A. Chen and R.V. Pappu, 
J. Phys. Chem. B {\bf 111}, 6469 (2007). 
 









\bibitem{Debye-Kleboth} P. Debye and K. Kleboth, 
  J. Chem. Phys. {\bf 42}, 3155 (1965).


 

\bibitem{gel} Yu. Kramarenko, I. Ya. Erukhimovich, 
and  A. R. Khokhlov,
Macromol. Theory and Simul. {\bf 11}, 
462 (2002). 


\bibitem{Born} M. Born, Z. Phys. {\bf 1}, 45 (1920). 
 

\bibitem{Hung}  
Le Quoc Hung, J. Electroanal. Chem. 
{\bf 115}, 159 (1980).

\bibitem{Osakai} T. Osakai and K. Ebina, 
J. Phys. Chem. B {\bf 102}, 5691 (1998). 


\bibitem{Hydrogen} K.V. Durme, H. Rahier, and B. van Mele, 
Macromolecules {\bf 38},  10155 (2005).



\bibitem{Seto1} K. Sadakane, H. Seto, H. Endo, and M. Kojima, 
J. Appl. Crystallogr., {\bf 40}, S527 (2007). 



\bibitem{polar1} E.L.  Eckfeldt and 
W.W. Lucasse, 
 {J. Phys. Chem.} {\bf 47}, 164 (1943).  
  
\bibitem{polar2} B.J. Hales, G.L. Bertrand, 
 and L.G. Hepler, 
 {J. Phys. Chem.} {\bf 70}, 3970 (1966). 


\bibitem{newly_found} 
V. Balevicius  and H.  Fuess, 
  Phys. Chem. Chem. Phys. {\bf{1}} ,1507 (1999). 









\end{references}
\end{document}